\documentclass[aps,prb,twocolumn,10pt]{revtex4-1}
\usepackage[T1]{fontenc}

\usepackage[pdftex]{graphicx}

\begin{document}

\title{Symmetry-dependent topological phase transitions in PbTe layers}

\author{Daniely Bassanezi}
\author{Ernesto Osvaldo Wrasse}
\affiliation{Universidade Tecnol\'ogica Federal do Paran\'a, Toledo, Paran\'a 85902-040, Brazil\\ }
\author{Tome M. Schmidt}
\affiliation{Instituto de F\'{\i}sica, Universidade Federal de Uberl\^andia,
 38400-902, Uberl\^andia, MG, Brazil}

\date{\today}

\begin{abstract}
By stacking PbTe layers there is a non-monotonic topological phase transition as a function of the number of monolayers.  Based on 
first principles calculations we find that the proper stacked crystal symmetry determines the topological nature of the slab. While a single
PbTe monolayer has a nontrivial phase, pressure can induce topological phase transition in bulk PbTe. Between these two limits, where
finite size effects are inherent, we verified that,
by applying an external pressure, odd stacking layers can be tuned easily 
to a topological phase, while even 
stacking keeps a larger band gap, avoiding band inversion. The quite distinct behavior for odd/even
layer is due to the symmetry of the finite stacking. Odd layers preserve the bulk symmorphic 
symmetry with strong surface hybridization, while even layers belong to a nonsymmorphic group symmetry. 
Nonsymmorphism induces extra degeneracies reducing the hybridization, thus
protecting band inversion, postponing topological phase transitions. 

\end{abstract}

                             

\maketitle

\section{\label{sec:level1}INTRODUCTION}
Topological insulators (TI) form a quantum phase of matter, where metallic states arise
on the borders and are protected by some symmetry. The first materials predicted in this class were protected
by time reversal symmetry \cite{kane}. Recently it has been proposed TIs protected by crystal symmetry \cite{fu}. 
This topological crystalline insulator (TCI) phase was predicted in a real material SnTe \cite{hsieh}, and promptly observed 
experimentally \cite{tanaka}, following many other IV-VI compounds
\cite{dziawa,PhysRevB.87.155105,xu,wojek,okada}. PbSe, PbTe and PbS three-dimensional (3D) bulk materials are trivial insulators,
but  pressure can be used to turn them in topological insulators \cite{barone,wan,PhysRevLett.115.086802}. On the other hand,
free standing monolayers composed by these materials are two-dimensional (2D) TCIs \cite{nl,Fu-NL-2015}.

The stacking of IV-VI monolayers formed by precursor bulk TCI materials have been largely studied
\cite{PhysRevB.87.235317,PhysRevB.90.045309,Liu-NMater-2014,Fu-NL-2015,PhysRevB.90.235114,
tanaka2013two,PhysRevB.91.081407}. 
Odd and even stacking of layers formed by TCI bulk materials show topological phase transitions and a non-monotonic
progression on their energy band gaps. In fact odd and even stacking can present different space crystal symmetry.
As this is the symmetry that protects the topological phase in these systems, odd/even can also present distinct topological phases. 
While odd stacking preserves the bulk symmorphic symmetry, even stacking present a nonsymmorphic symmetry. 
Nonsymmorphism by itself is already of great interest since it induces extra degeneracies, and it must be present in the 
lately proposed topological order Hourglass fermions \cite{Kane2015Nonsymm,Naturehourglass2016}.

Stacking of layers composed by a trivial precursor 3D bulk material can be interesting due to the connection of nano-size effects with
non trivial topological phases.
In this work we investigate topological properties due to finite size effects of stacked layers formed from the trivial bulk PbTe.
Nanostructure formed by PbTe is interesting to be investigated, since its 3D bulk is trivial,
while the monolayer is a 2D TCI system \cite{nl,Fu-NL-2015}. Based on electronic structure calculations we verify that for
odd stacking, due to the hybridization there is a compensation on the nano-size effects reducing the band gap.
On the other hand, for nonsymmorphic even stacking the symmetry-locked degeneracy keeps a larger band gap.
More interesting, we verified that a feasible pressure can support topological phase for
all odd stackings, while for nonsymmorphic even stacking, the topological phase can only be reached for a large number of
layers.

\section{\label{sec:level2}METHODOLOGY}

The PbTe nanostructures have been investigated using a density functional calculation (DFT)
as implemented in the VASP code \cite{vasp1,vasp2}. The 
exchange-correlation functional is described by the General Gradient Approximation \cite{pbe}. 
Interactions between electrons and ionic core are described by fully relativistic PAW 
pseudopotentials that include spin-orbit interactions \cite{paw}. Wave functions are expanded in 
orthogonal plane waves basis set with a energy cutoff of 450~eV. A Monkhorst-Pack \cite{monk} grid 
of $4\times4\times1$($4\times4\times4$) is employed to describe the BZ and obtain the charge density
of the layers (bulk). In all calculations a vacuum region of 12~ {\AA} in the perpendicular direction of 
the layers is used to reduce the interaction between periodic images.

\section{\label{sec:level3}RESULTS AND DISCUSSION}

Bulk PbTe, a rock salt structure, is an insulator material with small direct band gap (0.19 eV). The
spin orbit (SO) effect is quite strong in lead chalcogenides, particularly for PbTe our calculation energy band gap
is reduced from 0.83 eV to 0.11 eV, by including SO interactions. The smaller band gap obtained here as compared to the
experimental one is due to the DFT approach. The strong SO effect is not enough to invert the band gap, so 
bulk PbTe is a trivial insulator, where the valence band maximum (VBM) is composed mostly by $p$ orbitals of Te 
atoms and the conduction band minimum (CBM) by $p$ 
orbitals of Pb atoms. However, if pressure is applied to reduce the volume, we observe a band 
inversion that occurs
when PbTe is compressed to $91\%$ of the equilibrium volume $V_0$, making PbTe a nontrivial
insulator. Similar calculation has been performed by Barone {\it et. al.}, \cite{barone} where they obtained
also a topological phase transition due to pressure. The phase transition takes place due to the enhancement
of  $sp$ hybridization, pushing the $p$ states around the Fermi level provoking a band inversion.

\begin{figure}[h]
\includegraphics[width = 8.5cm,angle=0]{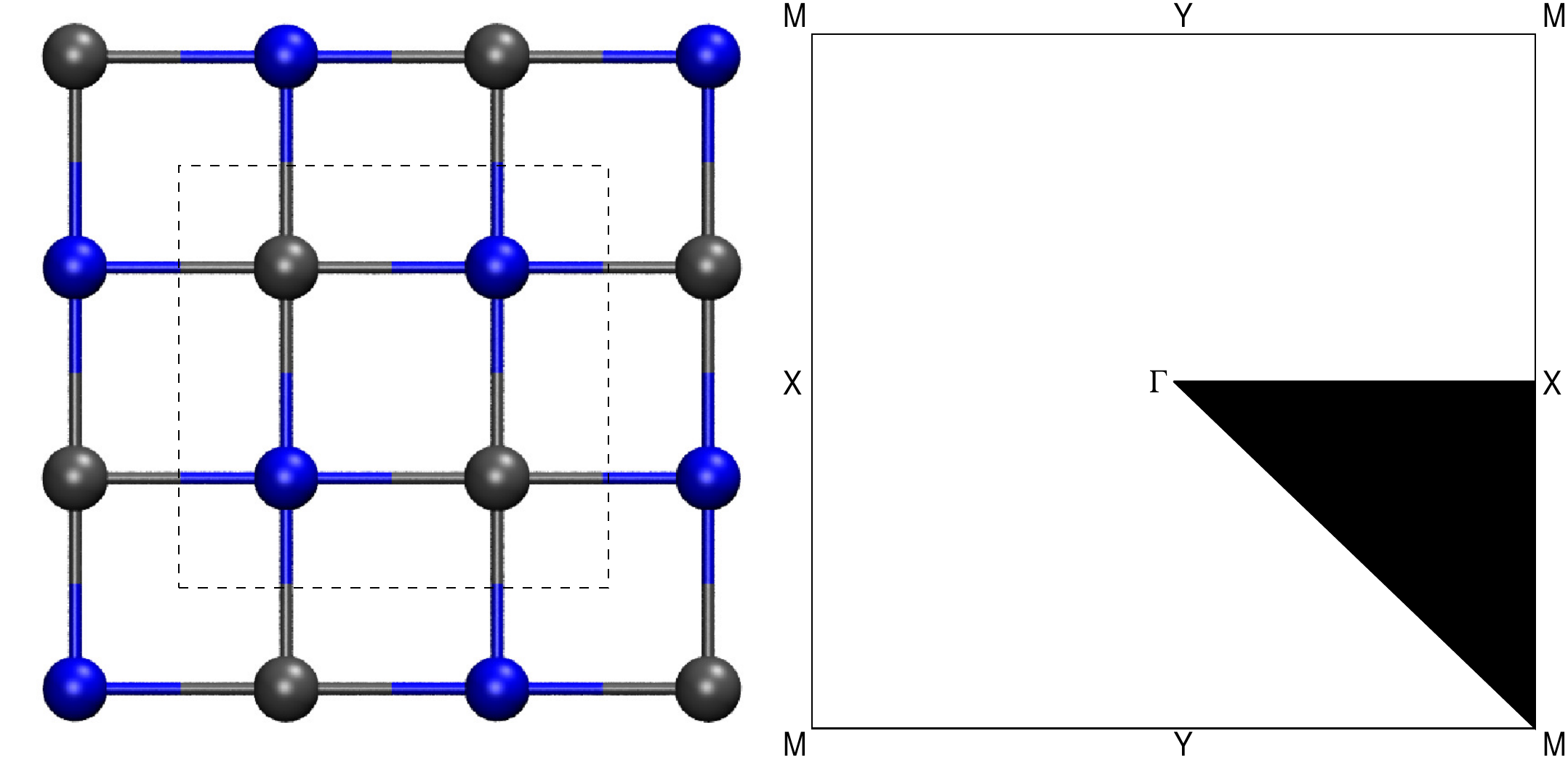}
\caption{\label{cell} Lattice structure and Brillouin Zone used here for PbTe monolayer. Blue spheres 
represent Pb atoms and grey spheres represent Te atoms. For this cell, the bulk L point is
projected into 2D M point.}
\end{figure}

\begin{figure}[htb]
\includegraphics[width = 8.5cm,angle=0]{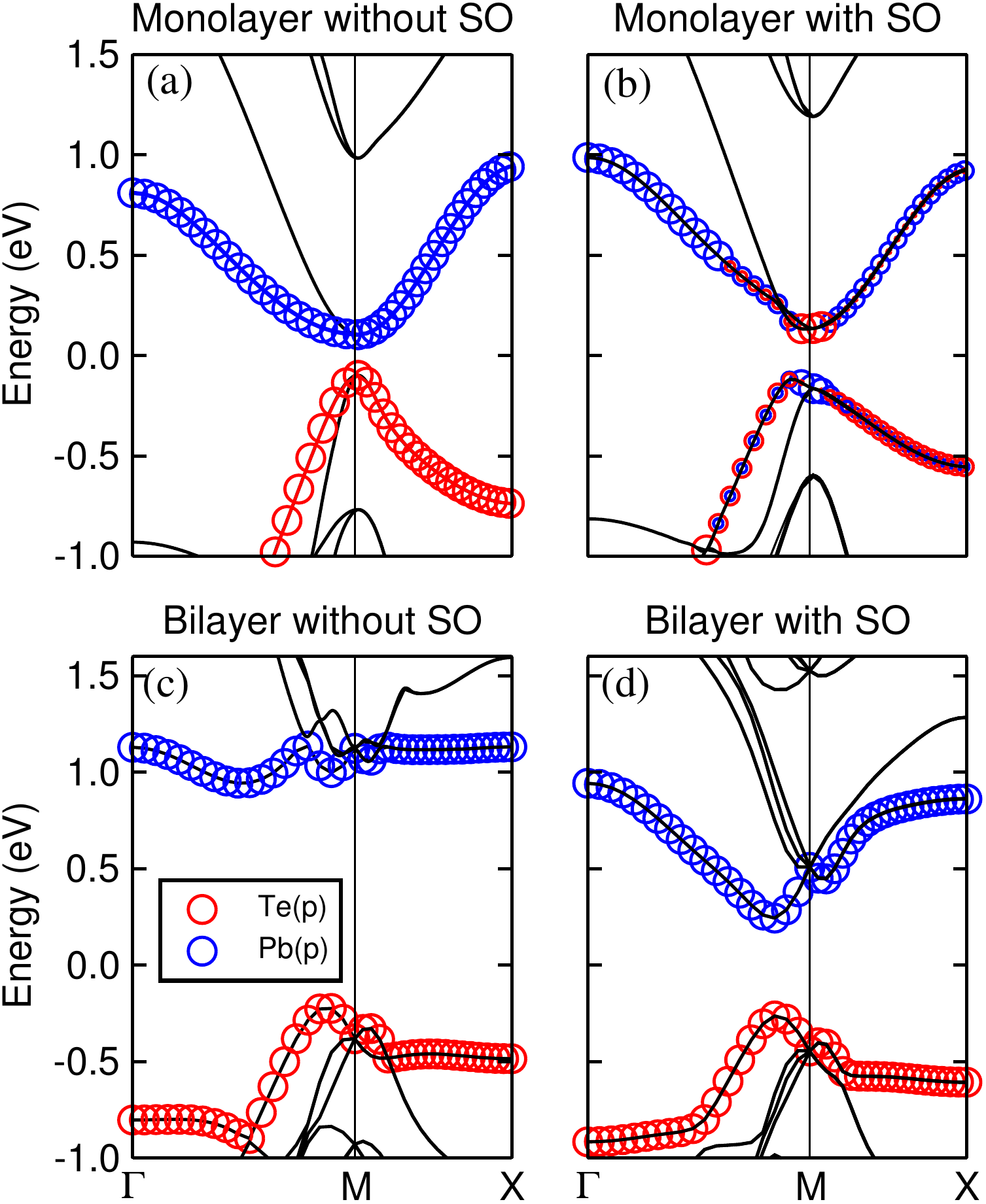}
\caption{\label{bilayer} Band structure of PbTe monolayer ((a) and (b)) and PbTe bilayer ((c) and 
(d)). Blue circles represent projection onto Pb $6p$ orbitals, while red circles projection onto Te $5p$ 
orbitals. The Fermi level is set to zero.}
\end{figure}

A nanoscale 2D system composed by a free standing monolayer of PbTe, showed in Fig.~\ref{cell}, presents a non-trivial
topological phase. Our calculated equilibrium lattice parameter for 
bulk PbTe is 6.56~{\AA}, and for the monolayer 6.36~{\AA}. This reduction in the lattice parameter 
strengthen the $sp$ hybridization, that, together with SO interactions, result in a band gap inversion 
without any external effect. In Fig.~\ref{bilayer}-ab we can see that the SO interactions put the Te $p$
orbitals in the conduction band and the Pb $p$ orbitals in the valence band, opposite characters as found in bulk PbTe.
Our results substantiate that PbTe monolayer is a 2D TCI material, in
agreement with previous calculations in 2D lead chalcogenides. \cite{nl,Fu-NL-2015}

\begin{figure}[hbt]
\includegraphics[width = 8.5cm,angle=0]{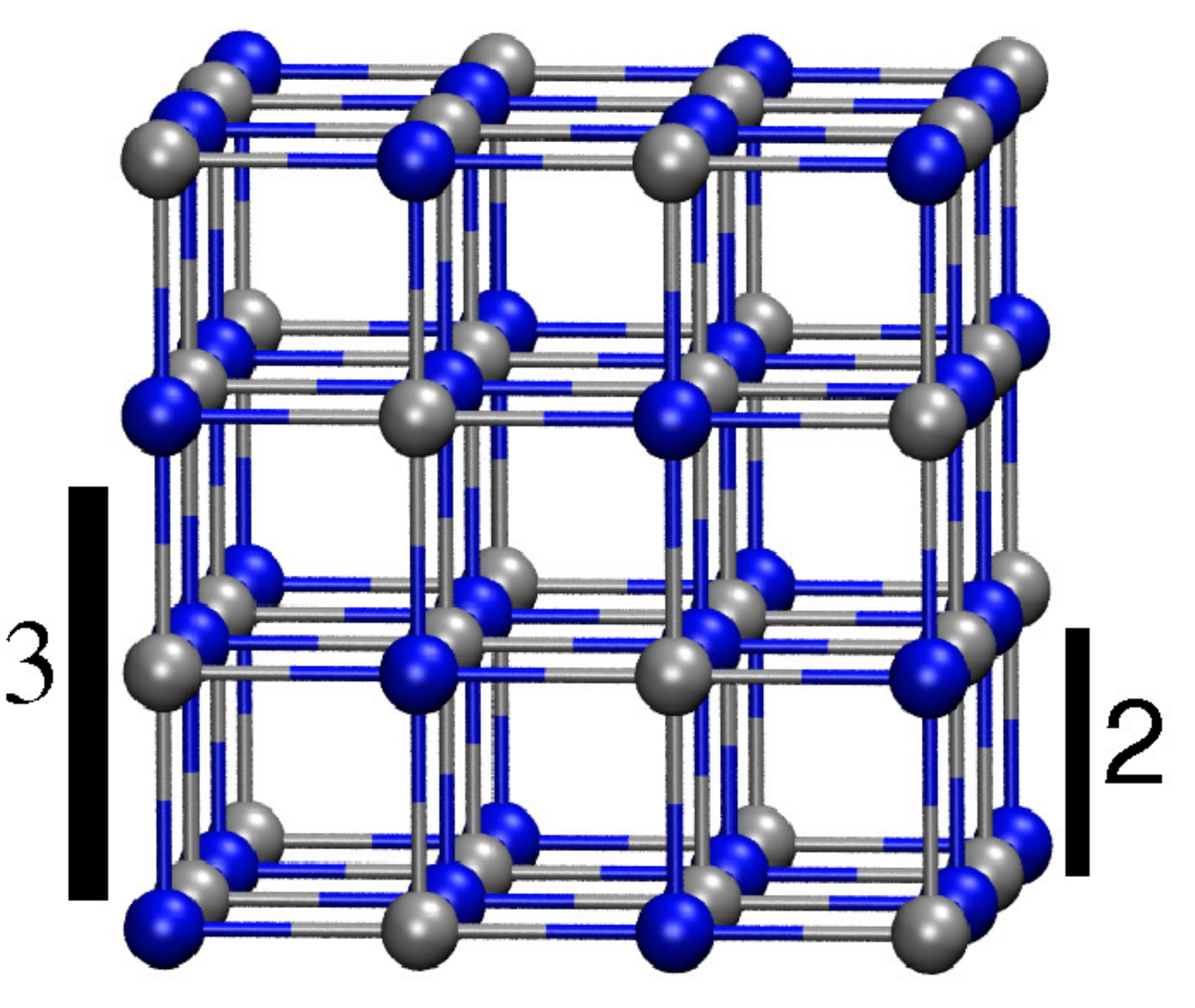}
\caption{\label{bulk} Lattice structure of PbTe few layers. For odd number of layers, like $N = 3$, the system is symmorphic, the
same symmetry as the bulk and monolayer of PbTe. For even number of layers, like $N = 2$, the system presents a 
nonsymmorphic symmetry. Blue and grey spheres represent Pb and Te atoms.}
\end{figure}

By stacking $N$ PbTe layers along the [001] direction as shown in Fig.~\ref{bulk} we can have odd or even number of layers.
For odd stacking the system preserves the same symmetry as the bulk and the monolayer PbTe. While for even number of layers the 
pure point group symmetry is not enough to describe the system, there are fractions of translation symmetry to be included. 
Odd/even stackings belong to different space group symmetries. In fact odd stackings present a 
nonsymmorphic  symmetry. The consequences of the nonsymmorphism are more interesting
on the electronic properties, where extra topological protections have been shown in 2D TCIs. \cite{prb3,PhysRevB.90.045309}

Let us starting with the nonsymmorphic bilayer ($N=2$), where the calculated equilibrium lattice parameter is 6.45~{\AA}, a value between the bulk and
the monolayer parameter. The bilayer presents larger band gap, 0.51~eV, as compared to the monolayer, 0.26~eV (see Fig.~\ref{bilayer}).
As the stacking enlarge the band gap, the SO interaction does not induce a band inversion, and the bilayer is a trivial insulator as
can be seen in Fig.~\ref{bilayer}-cd. At the 2D M point (L from the bulk) the electronic states are at the least
fourfold degenerate, due to the nonsymmorphic symmetry.

\begin{figure}[htb]
\includegraphics[width = 8.0cm,angle=0]{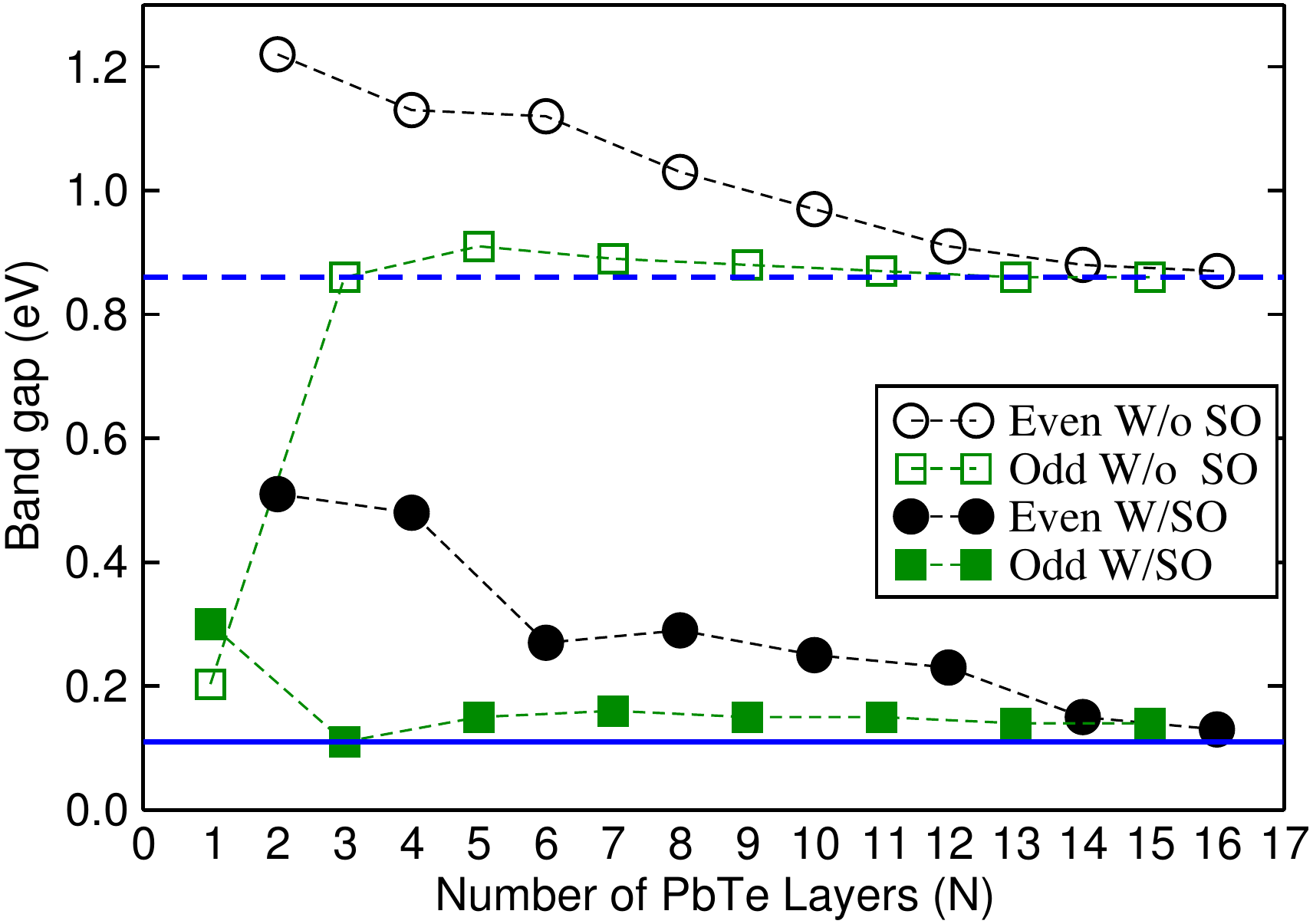}
\caption{\label{gap} Energy band gap as a function of number of PbTe layers $N$ along the [001] direction.
Open (filled) symbols represent band gap without (with) SO interaction. Circles (squares) are for even (odd) 
layer stacking. The full (dashed) horizontal line is the bulk PbTe band gap with (without) SO interaction.}
\end{figure}

For the symmorphic trilayer ($N=3$) we have a reduction on the band gap as compared to the bilayer, 0.11~eV and 0.86~eV with and without SO interaction, 
respectively. Those values are quite close to the bulk PbTe band gaps. However this 
reduction is not enough to change the bands, so trilayer PbTe is still within a trivial phase. By increasing the number $N$ 
of layers, as we can see from Fig.~\ref{gap}, the evolution of the energy band gaps are non-monotonic. While for symmorphic odd
stacking the band gap is almost constant, for nonsymmorphic even stacking the band gap varies with the number of layers.
Odd and even systems have to be analyzed separately, according to its spatial crystal symmetry.

Focusing on the symmorphic odd stackings, we can see a phase transition (with a gap closing) 
between 1 and 3 layers. For $N=1$ it is a TCI system, and for $N\ge3$ they are all trivial systems. A small oscillation of the band
gap can be seen for odd stacking in Fig.~\ref{gap}.
For the nonsymmorphic even stacking, the confinement effects affect strongly the electronic structure. For few layers the band gap 
is largely increased. We also observe a small oscillation of the band gap for even stacking, but no phase transition for 
these nonsymmorphic systems. The results are similar to those of Wan et al. \cite{wan} calculations for PbS stacking,
where odd/even present different trends on the band gap as a function of the number of layers.

\begin{figure}[htb]
\includegraphics[width = 8.5cm,angle=0]{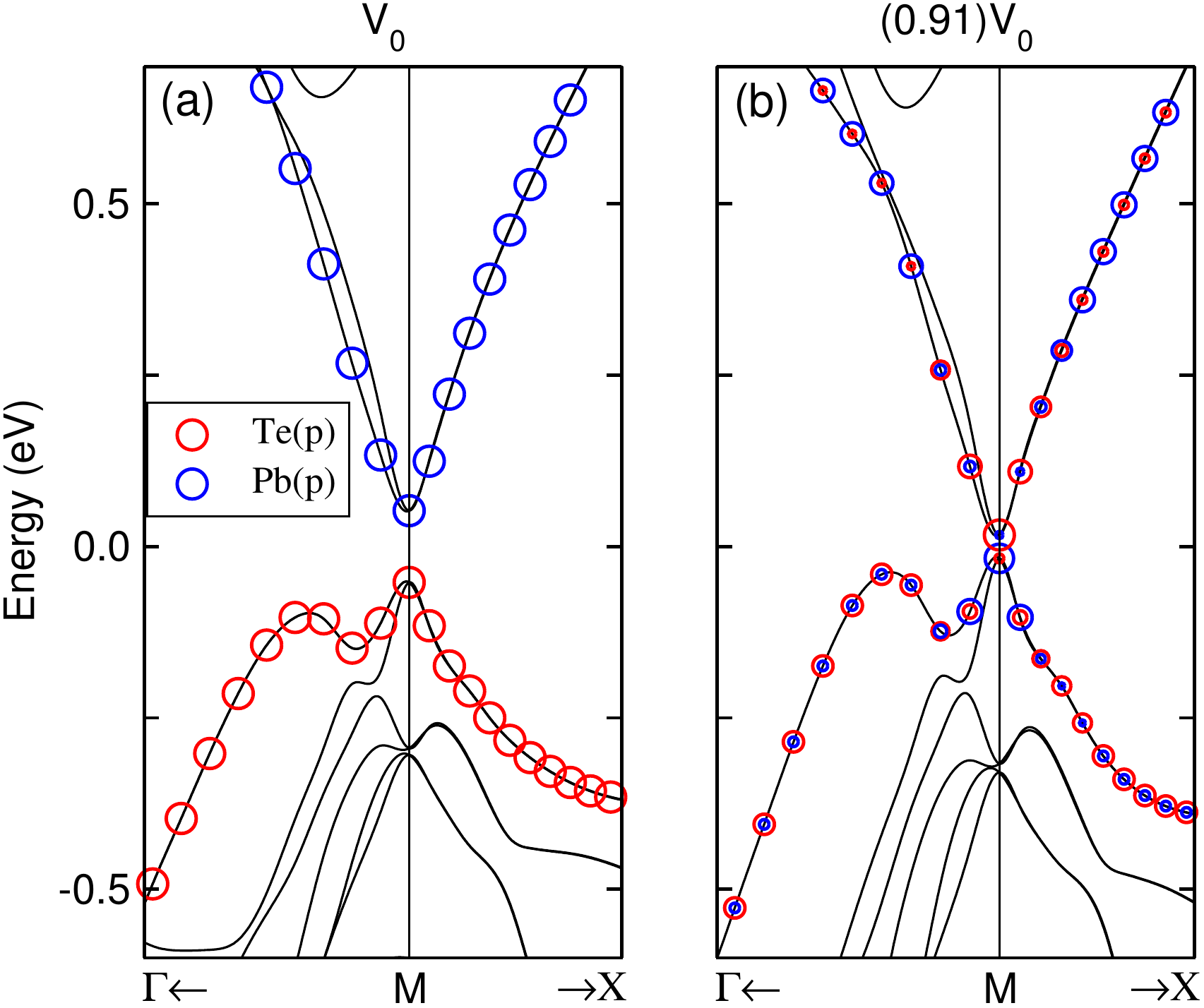}
\caption{\label{tri-press} Band structure of PbTe trilayer ($N=3$). In (a) is the band for optimized lattice 
parameter (volume $V_0$), and (b) is for an external pressure reducing the volume to $91\%$ of $V_0$.  
Blue circles represent projection onto Pb $6p$ orbitals, while red circles projection onto Te $5p$ 
orbitals. Fermi level is set to zero.}
\end{figure}

\begin{figure}[htb]
\includegraphics[width = 8.0cm,angle=0]{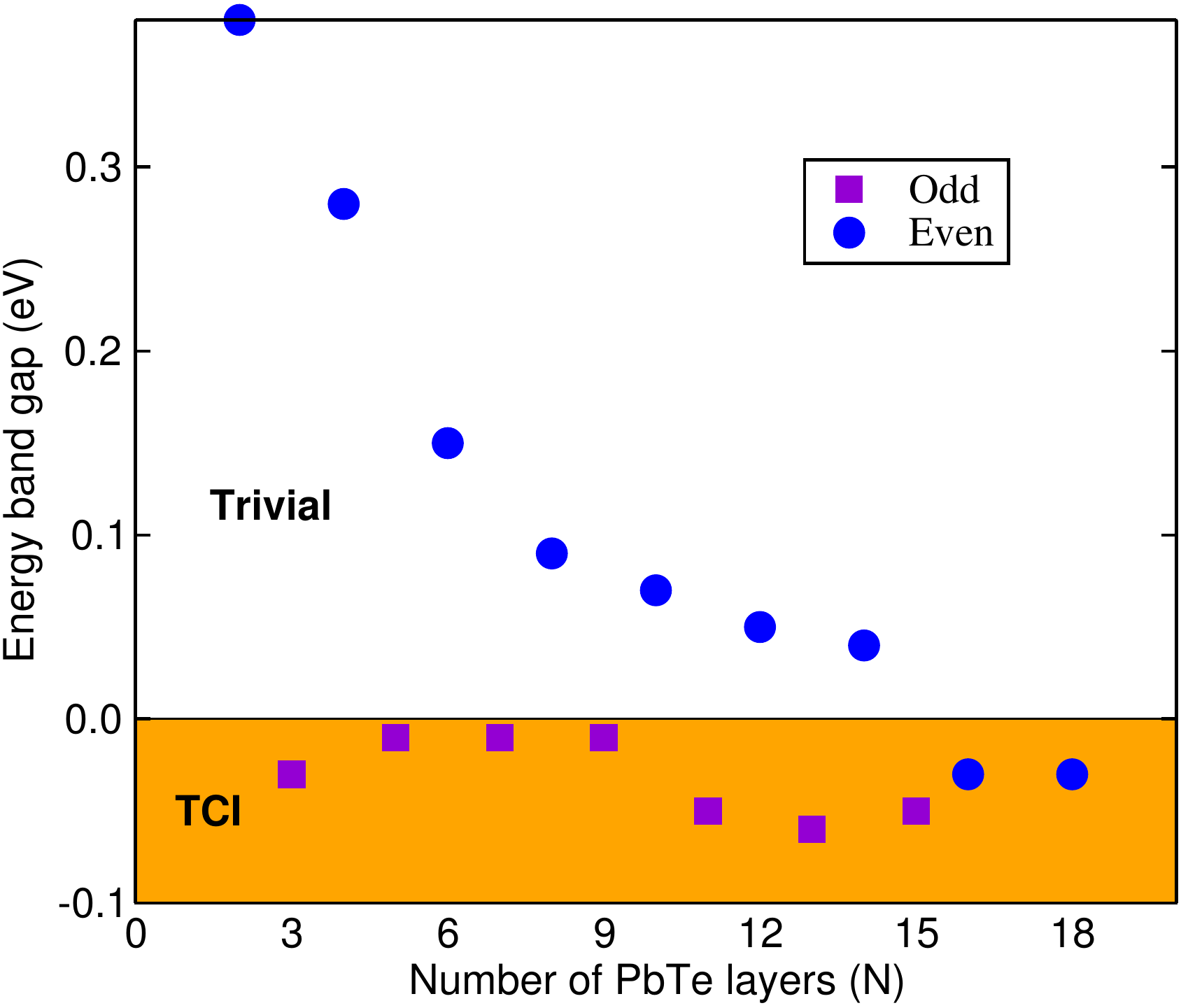}
\caption{\label{091} Energy band gap as a function of number of PbTe layers $N$ for an external pressure
reducing the volume to $91\%$ of the free standing layers. Negative values for the band gaps means 
topological non trivial phase.}
\end{figure}

It is interesting we analyze the topological properties by applying an external pressure on the PbTe layers, since bulk PbTe
changes it topological phase under pressure \cite{barone}. In particular for a trilayer, as shown in Fig.~\ref{tri-press}, we observe that
the fundamental band gap is already inverted for a reduction of the volume to $91\%$ of the free standing equilibrium volume $V_0$. 
In fact the topological phase transition occurs for a volume of $0.93V_0$, while for
bulk PbTe  this transition occurs for a volume of $0.91V_0$ \cite{barone}. For $N=5$ the trivial to topological phase occurs
for a volume of $0.92V_0$. And for all other $N\ge 7$ symmorphic odd PbTe stacking the band inversion takes place
for pressure of volume $0.91V_0$, the same as the bulk PbTe transition.  In  Fig.~\ref{091} we plot the energy band gap as 
a function of the number of layers for the critical volume $0.91V_0$. We observe that all symmorphic stackings belong to the TCI
phase. On the other hand, for nonsymmorphic even number of PbTe stacking, the trivial to topological phase only takes place 
for a large number of layers, $N\ge 16$. This postponing topological phase transition is due to the nonsymmorphic symmetry. 
Nonsymmorphism presents fractions of translation symmetry inducing extra degenerate states, hampering band inversions.

By comparing the evolution of the band gap for pressured systems of Fig.~\ref{091} with those of free standing of
Fig.~\ref{gap}, we see that the topological phase due to the compressed strain is reached for systems where the
non-pressured band gaps are close to the equilibrium volume of bulk PbTe band gap. As the confinement effects are
negligible for symmorphic odd stackings, keeping the band gaps
close to the PbTe bulk band gap (see Fig.~\ref{gap}), by reducing the volume to $0.91V_0$, all of they become TCI materials.
On the other hand, as the band gaps for nonsymmorphic even stackings are more sensitive with the number 
$N$ of PbTe layers,
the phase transition only occur for a large number of layers. Usually the effects of pressure in trivial 
systems enlarge the band gap, as observed for nonsymmorphic stacking with $N\le 14$ in Fig.~\ref{091}. If
the band gap is reduced due to the pressure, it may belong to a non-trivial phase, as is the case for $N\ge 16$.

\section{\label{sec:level4}CONCLUSIONS}
In summary we show that, while bulk 3D PbTe is a trivial material and a monolayer is a 2D TCI system, few layers of PbTe present
a non-monotonic evolution of the band gap, but they are still all trivial when free standing. Moreover by applying an external pressure,
odd stacking layers can be tuned easier to a topological phase than even stacking layers. As odd and even belong to
different spatial crystal symmetry, the confinement effects act different on each system. Few odd layers enhance the $sp$
hybridization leading to a faster band gap inversion. While the energy band gap for even nonsymmorphic stacking is quite
enlarged for few layers, avoiding a band inversion, even if pressure is applied. Based on electronic structure calculations,
we verified that all odd PbTe stackings become TCI systems when subject to a pressure of volume $0.91V_0$, the
same pressure as the bulk PbTe transition. On the other hand by keeping the same pressure, the topological phase transition
for even stacking only takes place for 16 layers or more. The difference on the topological phase transition for odd/even stacking
can be understood by the difference on the crystal symmetry. While odd layers preserve the bulk symmorphic symmetry,
even layers belong to a nonsymmorphic symmetry, which protects degenerate states, reducing the hybridization, thus avoiding 
band inversion and delaying the topological phase transition. 


\section{\label{sec:level1}ACKNOWLEDGMENTS}
The authors acknowledge the financial support from the Brazilian Agencies CNPq, CAPES, FAPEMIG, and 
ARAUCARIA and the computational facilities from CENAPAD.


\bibliography{PbTe-PRB}

\providecommand{\noopsort}[1]{}\providecommand{\singleletter}[1]{#1}%
\begin{thebibliography}{28}%
\makeatletter
\providecommand \@ifxundefined [1]{%
 \@ifx{#1\undefined}
}%
\providecommand \@ifnum [1]{%
 \ifnum #1\expandafter \@firstoftwo
 \else \expandafter \@secondoftwo
 \fi
}%
\providecommand \@ifx [1]{%
 \ifx #1\expandafter \@firstoftwo
 \else \expandafter \@secondoftwo
 \fi
}%
\providecommand \natexlab [1]{#1}%
\providecommand \enquote  [1]{``#1''}%
\providecommand \bibnamefont  [1]{#1}%
\providecommand \bibfnamefont [1]{#1}%
\providecommand \citenamefont [1]{#1}%
\providecommand \href@noop [0]{\@secondoftwo}%
\providecommand \href [0]{\begingroup \@sanitize@url \@href}%
\providecommand \@href[1]{\@@startlink{#1}\@@href}%
\providecommand \@@href[1]{\endgroup#1\@@endlink}%
\providecommand \@sanitize@url [0]{\catcode `\\12\catcode `\$12\catcode
  `\&12\catcode `\#12\catcode `\^12\catcode `\_12\catcode `\%12\relax}%
\providecommand \@@startlink[1]{}%
\providecommand \@@endlink[0]{}%
\providecommand \url  [0]{\begingroup\@sanitize@url \@url }%
\providecommand \@url [1]{\endgroup\@href {#1}{\urlprefix }}%
\providecommand \urlprefix  [0]{URL }%
\providecommand \Eprint [0]{\href }%
\providecommand \doibase [0]{http://dx.doi.org/}%
\providecommand \selectlanguage [0]{\@gobble}%
\providecommand \bibinfo  [0]{\@secondoftwo}%
\providecommand \bibfield  [0]{\@secondoftwo}%
\providecommand \translation [1]{[#1]}%
\providecommand \BibitemOpen [0]{}%
\providecommand \bibitemStop [0]{}%
\providecommand \bibitemNoStop [0]{.\EOS\space}%
\providecommand \EOS [0]{\spacefactor3000\relax}%
\providecommand \BibitemShut  [1]{\csname bibitem#1\endcsname}%
\let\auto@bib@innerbib\@empty
\bibitem [{\citenamefont {Hasan}\ and\ \citenamefont {Kane}(2010)}]{kane}%
  \BibitemOpen
  \bibfield  {author} {\bibinfo {author} {\bibfnamefont {M.~Z.}\ \bibnamefont
  {Hasan}}\ and\ \bibinfo {author} {\bibfnamefont {C.~L.}\ \bibnamefont
  {Kane}},\ }\href@noop {} {\bibfield  {journal} {\bibinfo  {journal} {Rev.
  Mod. Phys.}\ }\textbf {\bibinfo {volume} {82}},\ \bibinfo {pages} {3045}
  (\bibinfo {year} {2010})}\BibitemShut {NoStop}%
\bibitem [{\citenamefont {Fu}(2011)}]{fu}%
  \BibitemOpen
  \bibfield  {author} {\bibinfo {author} {\bibfnamefont {L.}~\bibnamefont
  {Fu}},\ }\href@noop {} {\bibfield  {journal} {\bibinfo  {journal} {Phys. Rev.
  Lett.}\ }\textbf {\bibinfo {volume} {106}},\ \bibinfo {pages} {106802}
  (\bibinfo {year} {2011})}\BibitemShut {NoStop}%
\bibitem [{\citenamefont {Hsieh}\ \emph {et~al.}(2012)\citenamefont {Hsieh},
  \citenamefont {Lin}, \citenamefont {Liu}, \citenamefont {Duan}, \citenamefont
  {Bansil},\ and\ \citenamefont {Fu}}]{hsieh}%
  \BibitemOpen
  \bibfield  {author} {\bibinfo {author} {\bibfnamefont {T.~H.}\ \bibnamefont
  {Hsieh}}, \bibinfo {author} {\bibfnamefont {H.}~\bibnamefont {Lin}}, \bibinfo
  {author} {\bibfnamefont {J.}~\bibnamefont {Liu}}, \bibinfo {author}
  {\bibfnamefont {W.}~\bibnamefont {Duan}}, \bibinfo {author} {\bibfnamefont
  {A.}~\bibnamefont {Bansil}}, \ and\ \bibinfo {author} {\bibfnamefont
  {L.}~\bibnamefont {Fu}},\ }\href@noop {} {\bibfield  {journal} {\bibinfo
  {journal} {Nat. Commun.}\ }\textbf {\bibinfo {volume} {3}},\ \bibinfo {pages}
  {982} (\bibinfo {year} {2012})}\BibitemShut {NoStop}%
\bibitem [{\citenamefont {Tanaka}\ \emph {et~al.}(2012)\citenamefont {Tanaka},
  \citenamefont {Ren}, \citenamefont {Sato}, \citenamefont {Nakayama},
  \citenamefont {Souma}, \citenamefont {Takahashi}, \citenamefont {Segawa},\
  and\ \citenamefont {Ando}}]{tanaka}%
  \BibitemOpen
  \bibfield  {author} {\bibinfo {author} {\bibfnamefont {Y.}~\bibnamefont
  {Tanaka}}, \bibinfo {author} {\bibfnamefont {Z.}~\bibnamefont {Ren}},
  \bibinfo {author} {\bibfnamefont {T.}~\bibnamefont {Sato}}, \bibinfo {author}
  {\bibfnamefont {K.}~\bibnamefont {Nakayama}}, \bibinfo {author}
  {\bibfnamefont {S.}~\bibnamefont {Souma}}, \bibinfo {author} {\bibfnamefont
  {T.}~\bibnamefont {Takahashi}}, \bibinfo {author} {\bibfnamefont
  {K.}~\bibnamefont {Segawa}}, \ and\ \bibinfo {author} {\bibfnamefont
  {Y.}~\bibnamefont {Ando}},\ }\href@noop {} {\bibfield  {journal} {\bibinfo
  {journal} {Nat. Phys.}\ }\textbf {\bibinfo {volume} {8}},\ \bibinfo {pages}
  {800} (\bibinfo {year} {2012})}\BibitemShut {NoStop}%
\bibitem [{\citenamefont {Dziawa}\ \emph {et~al.}(2012)\citenamefont {Dziawa},
  \citenamefont {Kowalski}, \citenamefont {Dybko}, \citenamefont {Buczko},
  \citenamefont {Szczerbakow}, \citenamefont {Szot}, \citenamefont
  {Lusakowska}, \citenamefont {Balasubramanian}, \citenamefont {Wojek},
  \citenamefont {Berntsen}, \citenamefont {Tjernberg},\ and\ \citenamefont
  {Story}}]{dziawa}%
  \BibitemOpen
  \bibfield  {author} {\bibinfo {author} {\bibfnamefont {P.}~\bibnamefont
  {Dziawa}}, \bibinfo {author} {\bibfnamefont {B.~J.}\ \bibnamefont
  {Kowalski}}, \bibinfo {author} {\bibfnamefont {K.}~\bibnamefont {Dybko}},
  \bibinfo {author} {\bibfnamefont {R.}~\bibnamefont {Buczko}}, \bibinfo
  {author} {\bibfnamefont {A.}~\bibnamefont {Szczerbakow}}, \bibinfo {author}
  {\bibfnamefont {M.}~\bibnamefont {Szot}}, \bibinfo {author} {\bibfnamefont
  {E.}~\bibnamefont {Lusakowska}}, \bibinfo {author} {\bibfnamefont
  {T.}~\bibnamefont {Balasubramanian}}, \bibinfo {author} {\bibfnamefont
  {B.~M.}\ \bibnamefont {Wojek}}, \bibinfo {author} {\bibfnamefont {M.~H.}\
  \bibnamefont {Berntsen}}, \bibinfo {author} {\bibfnamefont {O.}~\bibnamefont
  {Tjernberg}}, \ and\ \bibinfo {author} {\bibfnamefont {T.}~\bibnamefont
  {Story}},\ }\href@noop {} {\bibfield  {journal} {\bibinfo  {journal} {Nat.
  Mater.}\ }\textbf {\bibinfo {volume} {11}},\ \bibinfo {pages} {1023}
  (\bibinfo {year} {2012})}\BibitemShut {NoStop}%
\bibitem [{\citenamefont {Tanaka}\ \emph
  {et~al.}(2013{\natexlab{a}})\citenamefont {Tanaka}, \citenamefont {Sato},
  \citenamefont {Nakayama}, \citenamefont {Souma}, \citenamefont {Takahashi},
  \citenamefont {Ren}, \citenamefont {Novak}, \citenamefont {Segawa},\ and\
  \citenamefont {Ando}}]{PhysRevB.87.155105}%
  \BibitemOpen
  \bibfield  {author} {\bibinfo {author} {\bibfnamefont {Y.}~\bibnamefont
  {Tanaka}}, \bibinfo {author} {\bibfnamefont {T.}~\bibnamefont {Sato}},
  \bibinfo {author} {\bibfnamefont {K.}~\bibnamefont {Nakayama}}, \bibinfo
  {author} {\bibfnamefont {S.}~\bibnamefont {Souma}}, \bibinfo {author}
  {\bibfnamefont {T.}~\bibnamefont {Takahashi}}, \bibinfo {author}
  {\bibfnamefont {Z.}~\bibnamefont {Ren}}, \bibinfo {author} {\bibfnamefont
  {M.}~\bibnamefont {Novak}}, \bibinfo {author} {\bibfnamefont
  {K.}~\bibnamefont {Segawa}}, \ and\ \bibinfo {author} {\bibfnamefont
  {Y.}~\bibnamefont {Ando}},\ }\href {\doibase 10.1103/PhysRevB.87.155105}
  {\bibfield  {journal} {\bibinfo  {journal} {Phys. Rev. B}\ }\textbf {\bibinfo
  {volume} {87}},\ \bibinfo {pages} {155105} (\bibinfo {year}
  {2013}{\natexlab{a}})}\BibitemShut {NoStop}%
\bibitem [{\citenamefont {Xu}\ \emph {et~al.}(2012)\citenamefont {Xu},
  \citenamefont {Liu}, \citenamefont {Alidoust}, \citenamefont {Neupane},
  \citenamefont {Qian}, \citenamefont {Belopolski}, \citenamefont {Denlinger},
  \citenamefont {Wang}, \citenamefont {Lin}, \citenamefont {Wray},
  \citenamefont {Landolt}, \citenamefont {Slomski}, \citenamefont {Dil},
  \citenamefont {Marcinkova}, \citenamefont {Morosan}, \citenamefont {Gibson},
  \citenamefont {Sankar}, \citenamefont {Chou}, \citenamefont {Cava},
  \citenamefont {Bansil},\ and\ \citenamefont {Hasan}}]{xu}%
  \BibitemOpen
  \bibfield  {author} {\bibinfo {author} {\bibfnamefont {S.-Y.}\ \bibnamefont
  {Xu}}, \bibinfo {author} {\bibfnamefont {C.}~\bibnamefont {Liu}}, \bibinfo
  {author} {\bibfnamefont {N.}~\bibnamefont {Alidoust}}, \bibinfo {author}
  {\bibfnamefont {M.}~\bibnamefont {Neupane}}, \bibinfo {author} {\bibfnamefont
  {D.}~\bibnamefont {Qian}}, \bibinfo {author} {\bibfnamefont {I.}~\bibnamefont
  {Belopolski}}, \bibinfo {author} {\bibfnamefont {J.~D.}\ \bibnamefont
  {Denlinger}}, \bibinfo {author} {\bibfnamefont {Y.~J.}\ \bibnamefont {Wang}},
  \bibinfo {author} {\bibfnamefont {H.}~\bibnamefont {Lin}}, \bibinfo {author}
  {\bibfnamefont {L.~A.}\ \bibnamefont {Wray}}, \bibinfo {author}
  {\bibfnamefont {G.}~\bibnamefont {Landolt}}, \bibinfo {author} {\bibfnamefont
  {B.}~\bibnamefont {Slomski}}, \bibinfo {author} {\bibfnamefont {J.~H.}\
  \bibnamefont {Dil}}, \bibinfo {author} {\bibfnamefont {A.}~\bibnamefont
  {Marcinkova}}, \bibinfo {author} {\bibfnamefont {E.}~\bibnamefont {Morosan}},
  \bibinfo {author} {\bibfnamefont {Q.}~\bibnamefont {Gibson}}, \bibinfo
  {author} {\bibfnamefont {R.}~\bibnamefont {Sankar}}, \bibinfo {author}
  {\bibfnamefont {F.~C.}\ \bibnamefont {Chou}}, \bibinfo {author}
  {\bibfnamefont {R.~J.}\ \bibnamefont {Cava}}, \bibinfo {author}
  {\bibfnamefont {A.}~\bibnamefont {Bansil}}, \ and\ \bibinfo {author}
  {\bibfnamefont {M.~Z.}\ \bibnamefont {Hasan}},\ }\href@noop {} {\bibfield
  {journal} {\bibinfo  {journal} {Nat. Commun.}\ }\textbf {\bibinfo {volume}
  {3}},\ \bibinfo {pages} {1192} (\bibinfo {year} {2012})}\BibitemShut
  {NoStop}%
\bibitem [{\citenamefont {Wojek}\ \emph {et~al.}(2013)\citenamefont {Wojek},
  \citenamefont {Buczko}, \citenamefont {Safaei}, \citenamefont {Dziawa},
  \citenamefont {Kowalski}, \citenamefont {Berntsen}, \citenamefont
  {Balasubramanian}, \citenamefont {T}, \citenamefont {A}, \citenamefont
  {Kacman}, \citenamefont {Story},\ and\ \citenamefont {Tjernber}}]{wojek}%
  \BibitemOpen
  \bibfield  {author} {\bibinfo {author} {\bibfnamefont {B.~M.}\ \bibnamefont
  {Wojek}}, \bibinfo {author} {\bibfnamefont {R.}~\bibnamefont {Buczko}},
  \bibinfo {author} {\bibfnamefont {S.}~\bibnamefont {Safaei}}, \bibinfo
  {author} {\bibfnamefont {P.}~\bibnamefont {Dziawa}}, \bibinfo {author}
  {\bibfnamefont {B.~J.}\ \bibnamefont {Kowalski}}, \bibinfo {author}
  {\bibfnamefont {M.~H.}\ \bibnamefont {Berntsen}}, \bibinfo {author}
  {\bibfnamefont {T.}~\bibnamefont {Balasubramanian}}, \bibinfo {author}
  {\bibfnamefont {L.}~\bibnamefont {T}}, \bibinfo {author} {\bibfnamefont
  {S.}~\bibnamefont {A}}, \bibinfo {author} {\bibfnamefont {P.}~\bibnamefont
  {Kacman}}, \bibinfo {author} {\bibfnamefont {T.}~\bibnamefont {Story}}, \
  and\ \bibinfo {author} {\bibfnamefont {O.}~\bibnamefont {Tjernber}},\
  }\href@noop {} {\bibfield  {journal} {\bibinfo  {journal} {Phys. Rev. B}\
  }\textbf {\bibinfo {volume} {87}},\ \bibinfo {pages} {115106} (\bibinfo
  {year} {2013})}\BibitemShut {NoStop}%
\bibitem [{\citenamefont {Okada}\ \emph {et~al.}(2013)\citenamefont {Okada},
  \citenamefont {Serbyn}, \citenamefont {Lin}, \citenamefont {Walkup},
  \citenamefont {Zhou}, \citenamefont {Dhital}, \citenamefont {Neupane},
  \citenamefont {Xu}, \citenamefont {Wang}, \citenamefont {Sankar},
  \citenamefont {Chou}, \citenamefont {Bansil}, \citenamefont {Hasan},
  \citenamefont {Wilson}, \citenamefont {Fu},\ and\ \citenamefont
  {Madhavan}}]{okada}%
  \BibitemOpen
  \bibfield  {author} {\bibinfo {author} {\bibfnamefont {Y.}~\bibnamefont
  {Okada}}, \bibinfo {author} {\bibfnamefont {M.}~\bibnamefont {Serbyn}},
  \bibinfo {author} {\bibfnamefont {H.}~\bibnamefont {Lin}}, \bibinfo {author}
  {\bibfnamefont {D.}~\bibnamefont {Walkup}}, \bibinfo {author} {\bibfnamefont
  {W.}~\bibnamefont {Zhou}}, \bibinfo {author} {\bibfnamefont {C.}~\bibnamefont
  {Dhital}}, \bibinfo {author} {\bibfnamefont {M.}~\bibnamefont {Neupane}},
  \bibinfo {author} {\bibfnamefont {S.}~\bibnamefont {Xu}}, \bibinfo {author}
  {\bibfnamefont {Y.~J.}\ \bibnamefont {Wang}}, \bibinfo {author}
  {\bibfnamefont {R.}~\bibnamefont {Sankar}}, \bibinfo {author} {\bibfnamefont
  {F.}~\bibnamefont {Chou}}, \bibinfo {author} {\bibfnamefont {A.}~\bibnamefont
  {Bansil}}, \bibinfo {author} {\bibfnamefont {M.~Z.}\ \bibnamefont {Hasan}},
  \bibinfo {author} {\bibfnamefont {S.~D.}\ \bibnamefont {Wilson}}, \bibinfo
  {author} {\bibfnamefont {L.}~\bibnamefont {Fu}}, \ and\ \bibinfo {author}
  {\bibfnamefont {V.}~\bibnamefont {Madhavan}},\ }\href@noop {} {\bibfield
  {journal} {\bibinfo  {journal} {Science}\ }\textbf {\bibinfo {volume}
  {341}},\ \bibinfo {pages} {1496} (\bibinfo {year} {2013})}\BibitemShut
  {NoStop}%
\bibitem [{\citenamefont {Barone}\ \emph {et~al.}(2013)\citenamefont {Barone},
  \citenamefont {Rauch}, \citenamefont {Sante}, \citenamefont {Henk},
  \citenamefont {Mertig},\ and\ \citenamefont {Picozzi}}]{barone}%
  \BibitemOpen
  \bibfield  {author} {\bibinfo {author} {\bibfnamefont {P.}~\bibnamefont
  {Barone}}, \bibinfo {author} {\bibfnamefont {T.}~\bibnamefont {Rauch}},
  \bibinfo {author} {\bibfnamefont {D.~D.}\ \bibnamefont {Sante}}, \bibinfo
  {author} {\bibfnamefont {J.}~\bibnamefont {Henk}}, \bibinfo {author}
  {\bibfnamefont {I.}~\bibnamefont {Mertig}}, \ and\ \bibinfo {author}
  {\bibfnamefont {S.}~\bibnamefont {Picozzi}},\ }\href@noop {} {\bibfield
  {journal} {\bibinfo  {journal} {Phys. Rev. B}\ }\textbf {\bibinfo {volume}
  {88}},\ \bibinfo {pages} {045207} (\bibinfo {year} {2013})}\BibitemShut
  {NoStop}%
\bibitem [{\citenamefont {Wan}\ \emph {et~al.}(2017)\citenamefont {Wan},
  \citenamefont {Yao}, \citenamefont {Sun}, \citenamefont {Liu},\ and\
  \citenamefont {Zhang}}]{wan}%
  \BibitemOpen
  \bibfield  {author} {\bibinfo {author} {\bibfnamefont {W.}~\bibnamefont
  {Wan}}, \bibinfo {author} {\bibfnamefont {Y.}~\bibnamefont {Yao}}, \bibinfo
  {author} {\bibfnamefont {L.}~\bibnamefont {Sun}}, \bibinfo {author}
  {\bibfnamefont {C.-C.}\ \bibnamefont {Liu}}, \ and\ \bibinfo {author}
  {\bibfnamefont {F.}~\bibnamefont {Zhang}},\ }\href@noop {} {\bibfield
  {journal} {\bibinfo  {journal} {Advanced Mat.}\ }\textbf {\bibinfo {volume}
  {29}},\ \bibinfo {pages} {1604788} (\bibinfo {year} {2017})}\BibitemShut
  {NoStop}%
\bibitem [{\citenamefont {Kim}\ \emph {et~al.}(2015)\citenamefont {Kim},
  \citenamefont {Kane}, \citenamefont {Mele},\ and\ \citenamefont
  {Rappe}}]{PhysRevLett.115.086802}%
  \BibitemOpen
  \bibfield  {author} {\bibinfo {author} {\bibfnamefont {Y.}~\bibnamefont
  {Kim}}, \bibinfo {author} {\bibfnamefont {C.~L.}\ \bibnamefont {Kane}},
  \bibinfo {author} {\bibfnamefont {E.~J.}\ \bibnamefont {Mele}}, \ and\
  \bibinfo {author} {\bibfnamefont {A.~M.}\ \bibnamefont {Rappe}},\ }\href
  {\doibase 10.1103/PhysRevLett.115.086802} {\bibfield  {journal} {\bibinfo
  {journal} {Phys. Rev. Lett.}\ }\textbf {\bibinfo {volume} {115}},\ \bibinfo
  {pages} {086802} (\bibinfo {year} {2015})}\BibitemShut {NoStop}%
\bibitem [{\citenamefont {Wrasse}\ and\ \citenamefont {Schmidt}(2014)}]{nl}%
  \BibitemOpen
  \bibfield  {author} {\bibinfo {author} {\bibfnamefont {E.~O.}\ \bibnamefont
  {Wrasse}}\ and\ \bibinfo {author} {\bibfnamefont {T.~M.}\ \bibnamefont
  {Schmidt}},\ }\href@noop {} {\bibfield  {journal} {\bibinfo  {journal} {Nano
  Lett.}\ }\textbf {\bibinfo {volume} {14}},\ \bibinfo {pages} {5717} (\bibinfo
  {year} {2014})}\BibitemShut {NoStop}%
\bibitem [{\citenamefont {Liu}\ \emph {et~al.}(2015)\citenamefont {Liu},
  \citenamefont {Qian},\ and\ \citenamefont {Fu}}]{Fu-NL-2015}%
  \BibitemOpen
  \bibfield  {author} {\bibinfo {author} {\bibfnamefont {J.}~\bibnamefont
  {Liu}}, \bibinfo {author} {\bibfnamefont {X.}~\bibnamefont {Qian}}, \ and\
  \bibinfo {author} {\bibfnamefont {L.}~\bibnamefont {Fu}},\ }\href@noop {}
  {\bibfield  {journal} {\bibinfo  {journal} {Nano letters}\ }\textbf {\bibinfo
  {volume} {15}},\ \bibinfo {pages} {2657} (\bibinfo {year}
  {2015})}\BibitemShut {NoStop}%
\bibitem [{\citenamefont {Wang}\ \emph {et~al.}(2013)\citenamefont {Wang},
  \citenamefont {Tsai}, \citenamefont {Lin}, \citenamefont {Xu}, \citenamefont
  {Neupane}, \citenamefont {Hasan},\ and\ \citenamefont
  {Bansil}}]{PhysRevB.87.235317}%
  \BibitemOpen
  \bibfield  {author} {\bibinfo {author} {\bibfnamefont {Y.~J.}\ \bibnamefont
  {Wang}}, \bibinfo {author} {\bibfnamefont {W.-F.}\ \bibnamefont {Tsai}},
  \bibinfo {author} {\bibfnamefont {H.}~\bibnamefont {Lin}}, \bibinfo {author}
  {\bibfnamefont {S.-Y.}\ \bibnamefont {Xu}}, \bibinfo {author} {\bibfnamefont
  {M.}~\bibnamefont {Neupane}}, \bibinfo {author} {\bibfnamefont {M.~Z.}\
  \bibnamefont {Hasan}}, \ and\ \bibinfo {author} {\bibfnamefont
  {A.}~\bibnamefont {Bansil}},\ }\href {\doibase 10.1103/PhysRevB.87.235317}
  {\bibfield  {journal} {\bibinfo  {journal} {Phys. Rev. B}\ }\textbf {\bibinfo
  {volume} {87}},\ \bibinfo {pages} {235317} (\bibinfo {year}
  {2013})}\BibitemShut {NoStop}%
\bibitem [{\citenamefont {Ozawa}\ \emph {et~al.}(2014)\citenamefont {Ozawa},
  \citenamefont {Yamakage}, \citenamefont {Sato},\ and\ \citenamefont
  {Tanaka}}]{PhysRevB.90.045309}%
  \BibitemOpen
  \bibfield  {author} {\bibinfo {author} {\bibfnamefont {H.}~\bibnamefont
  {Ozawa}}, \bibinfo {author} {\bibfnamefont {A.}~\bibnamefont {Yamakage}},
  \bibinfo {author} {\bibfnamefont {M.}~\bibnamefont {Sato}}, \ and\ \bibinfo
  {author} {\bibfnamefont {Y.}~\bibnamefont {Tanaka}},\ }\href {\doibase
  10.1103/PhysRevB.90.045309} {\bibfield  {journal} {\bibinfo  {journal} {Phys.
  Rev. B}\ }\textbf {\bibinfo {volume} {90}},\ \bibinfo {pages} {045309}
  (\bibinfo {year} {2014})}\BibitemShut {NoStop}%
\bibitem [{\citenamefont {Liu}\ \emph {et~al.}(2014)\citenamefont {Liu},
  \citenamefont {Hsieh}, \citenamefont {Wei}, \citenamefont {Duan},
  \citenamefont {Moodera},\ and\ \citenamefont {Fu}}]{Liu-NMater-2014}%
  \BibitemOpen
  \bibfield  {author} {\bibinfo {author} {\bibfnamefont {J.}~\bibnamefont
  {Liu}}, \bibinfo {author} {\bibfnamefont {T.~H.}\ \bibnamefont {Hsieh}},
  \bibinfo {author} {\bibfnamefont {P.}~\bibnamefont {Wei}}, \bibinfo {author}
  {\bibfnamefont {W.}~\bibnamefont {Duan}}, \bibinfo {author} {\bibfnamefont
  {J.}~\bibnamefont {Moodera}}, \ and\ \bibinfo {author} {\bibfnamefont
  {L.}~\bibnamefont {Fu}},\ }\href@noop {} {\bibfield  {journal} {\bibinfo
  {journal} {Nature Materials}\ }\textbf {\bibinfo {volume} {13}},\ \bibinfo
  {pages} {178} (\bibinfo {year} {2014})}\BibitemShut {NoStop}%
\bibitem [{\citenamefont {Shi}\ \emph {et~al.}(2014)\citenamefont {Shi},
  \citenamefont {Wu}, \citenamefont {Zhang},\ and\ \citenamefont
  {Feng}}]{PhysRevB.90.235114}%
  \BibitemOpen
  \bibfield  {author} {\bibinfo {author} {\bibfnamefont {Y.}~\bibnamefont
  {Shi}}, \bibinfo {author} {\bibfnamefont {M.}~\bibnamefont {Wu}}, \bibinfo
  {author} {\bibfnamefont {F.}~\bibnamefont {Zhang}}, \ and\ \bibinfo {author}
  {\bibfnamefont {J.}~\bibnamefont {Feng}},\ }\href {\doibase
  10.1103/PhysRevB.90.235114} {\bibfield  {journal} {\bibinfo  {journal} {Phys.
  Rev. B}\ }\textbf {\bibinfo {volume} {90}},\ \bibinfo {pages} {235114}
  (\bibinfo {year} {2014})}\BibitemShut {NoStop}%
\bibitem [{\citenamefont {Tanaka}\ \emph
  {et~al.}(2013{\natexlab{b}})\citenamefont {Tanaka}, \citenamefont {Shoman},
  \citenamefont {Nakayama}, \citenamefont {Souma}, \citenamefont {Sato},
  \citenamefont {Takahashi}, \citenamefont {Novak}, \citenamefont {Segawa},\
  and\ \citenamefont {Ando}}]{tanaka2013two}%
  \BibitemOpen
  \bibfield  {author} {\bibinfo {author} {\bibfnamefont {Y.}~\bibnamefont
  {Tanaka}}, \bibinfo {author} {\bibfnamefont {T.}~\bibnamefont {Shoman}},
  \bibinfo {author} {\bibfnamefont {K.}~\bibnamefont {Nakayama}}, \bibinfo
  {author} {\bibfnamefont {S.}~\bibnamefont {Souma}}, \bibinfo {author}
  {\bibfnamefont {T.}~\bibnamefont {Sato}}, \bibinfo {author} {\bibfnamefont
  {T.}~\bibnamefont {Takahashi}}, \bibinfo {author} {\bibfnamefont
  {M.}~\bibnamefont {Novak}}, \bibinfo {author} {\bibfnamefont
  {K.}~\bibnamefont {Segawa}}, \ and\ \bibinfo {author} {\bibfnamefont
  {Y.}~\bibnamefont {Ando}},\ }\href@noop {} {\bibfield  {journal} {\bibinfo
  {journal} {Physical Review B}\ }\textbf {\bibinfo {volume} {88}},\ \bibinfo
  {pages} {235126} (\bibinfo {year} {2013}{\natexlab{b}})}\BibitemShut
  {NoStop}%
\bibitem [{\citenamefont {Liu}\ and\ \citenamefont
  {Fu}(2015)}]{PhysRevB.91.081407}%
  \BibitemOpen
  \bibfield  {author} {\bibinfo {author} {\bibfnamefont {J.}~\bibnamefont
  {Liu}}\ and\ \bibinfo {author} {\bibfnamefont {L.}~\bibnamefont {Fu}},\
  }\href {\doibase 10.1103/PhysRevB.91.081407} {\bibfield  {journal} {\bibinfo
  {journal} {Phys. Rev. B}\ }\textbf {\bibinfo {volume} {91}},\ \bibinfo
  {pages} {081407} (\bibinfo {year} {2015})}\BibitemShut {NoStop}%
\bibitem [{\citenamefont {Young}\ and\ \citenamefont
  {Kane}(2015)}]{Kane2015Nonsymm}%
  \BibitemOpen
  \bibfield  {author} {\bibinfo {author} {\bibfnamefont {S.~M.}\ \bibnamefont
  {Young}}\ and\ \bibinfo {author} {\bibfnamefont {C.~L.}\ \bibnamefont
  {Kane}},\ }\href {\doibase 10.1103/PhysRevLett.115.126803} {\bibfield
  {journal} {\bibinfo  {journal} {Phys. Rev. Lett.}\ }\textbf {\bibinfo
  {volume} {115}},\ \bibinfo {pages} {126803} (\bibinfo {year}
  {2015})}\BibitemShut {NoStop}%
\bibitem [{\citenamefont {Wang}\ \emph {et~al.}(2016)\citenamefont {Wang},
  \citenamefont {Alexandradinata}, \citenamefont {Cava},\ and\ \citenamefont
  {Bernevig}}]{Naturehourglass2016}%
  \BibitemOpen
  \bibfield  {author} {\bibinfo {author} {\bibfnamefont {Z.}~\bibnamefont
  {Wang}}, \bibinfo {author} {\bibfnamefont {A.}~\bibnamefont
  {Alexandradinata}}, \bibinfo {author} {\bibfnamefont {R.~J.}\ \bibnamefont
  {Cava}}, \ and\ \bibinfo {author} {\bibfnamefont {B.~A.}\ \bibnamefont
  {Bernevig}},\ }\href {\doibase 10.1038/nature17410} {\bibfield  {journal}
  {\bibinfo  {journal} {Nature}\ }\textbf {\bibinfo {volume} {532}},\ \bibinfo
  {pages} {189} (\bibinfo {year} {2016})}\BibitemShut {NoStop}%
\bibitem [{\citenamefont {Kresse}\ and\ \citenamefont
  {Furthmuller}(1996{\natexlab{a}})}]{vasp1}%
  \BibitemOpen
  \bibfield  {author} {\bibinfo {author} {\bibfnamefont {G.}~\bibnamefont
  {Kresse}}\ and\ \bibinfo {author} {\bibfnamefont {J.}~\bibnamefont
  {Furthmuller}},\ }\href@noop {} {\bibfield  {journal} {\bibinfo  {journal}
  {Phys. Rev. B}\ }\textbf {\bibinfo {volume} {54}},\ \bibinfo {pages} {11169}
  (\bibinfo {year} {1996}{\natexlab{a}})}\BibitemShut {NoStop}%
\bibitem [{\citenamefont {Kresse}\ and\ \citenamefont
  {Furthmuller}(1996{\natexlab{b}})}]{vasp2}%
  \BibitemOpen
  \bibfield  {author} {\bibinfo {author} {\bibfnamefont {G.}~\bibnamefont
  {Kresse}}\ and\ \bibinfo {author} {\bibfnamefont {J.}~\bibnamefont
  {Furthmuller}},\ }\href@noop {} {\bibfield  {journal} {\bibinfo  {journal}
  {Comput. Mater. Sci.}\ }\textbf {\bibinfo {volume} {6}},\ \bibinfo {pages}
  {15} (\bibinfo {year} {1996}{\natexlab{b}})}\BibitemShut {NoStop}%
\bibitem [{\citenamefont {Perdew}\ \emph {et~al.}(1996)\citenamefont {Perdew},
  \citenamefont {Burke},\ and\ \citenamefont {Ernzerhof}}]{pbe}%
  \BibitemOpen
  \bibfield  {author} {\bibinfo {author} {\bibfnamefont {J.~P.}\ \bibnamefont
  {Perdew}}, \bibinfo {author} {\bibfnamefont {K.}~\bibnamefont {Burke}}, \
  and\ \bibinfo {author} {\bibfnamefont {M.}~\bibnamefont {Ernzerhof}},\
  }\href@noop {} {\bibfield  {journal} {\bibinfo  {journal} {Phys. Rev. Lett.}\
  }\textbf {\bibinfo {volume} {77}},\ \bibinfo {pages} {3865} (\bibinfo {year}
  {1996})}\BibitemShut {NoStop}%
\bibitem [{\citenamefont {Kresse}\ and\ \citenamefont {Joubert}(1999)}]{paw}%
  \BibitemOpen
  \bibfield  {author} {\bibinfo {author} {\bibfnamefont {G.}~\bibnamefont
  {Kresse}}\ and\ \bibinfo {author} {\bibfnamefont {D.}~\bibnamefont
  {Joubert}},\ }\href@noop {} {\bibfield  {journal} {\bibinfo  {journal} {Phys.
  Rev. B}\ }\textbf {\bibinfo {volume} {59}},\ \bibinfo {pages} {1758}
  (\bibinfo {year} {1999})}\BibitemShut {NoStop}%
\bibitem [{\citenamefont {Monkhorst}\ and\ \citenamefont {Pack}(1976)}]{monk}%
  \BibitemOpen
  \bibfield  {author} {\bibinfo {author} {\bibfnamefont {H.~J.}\ \bibnamefont
  {Monkhorst}}\ and\ \bibinfo {author} {\bibfnamefont {J.~D.}\ \bibnamefont
  {Pack}},\ }\href@noop {} {\bibfield  {journal} {\bibinfo  {journal} {Phys.
  Rev. B}\ }\textbf {\bibinfo {volume} {13}},\ \bibinfo {pages} {5188}
  (\bibinfo {year} {1976})}\BibitemShut {NoStop}%
\bibitem [{\citenamefont {Araujo}\ \emph {et~al.}(2016)\citenamefont {Araujo},
  \citenamefont {Wrasse}, \citenamefont {Ferreira},\ and\ \citenamefont
  {Schmidt}}]{prb3}%
  \BibitemOpen
  \bibfield  {author} {\bibinfo {author} {\bibfnamefont {A.~L.}\ \bibnamefont
  {Araujo}}, \bibinfo {author} {\bibfnamefont {E.~O.}\ \bibnamefont {Wrasse}},
  \bibinfo {author} {\bibfnamefont {G.~J.}\ \bibnamefont {Ferreira}}, \ and\
  \bibinfo {author} {\bibfnamefont {T.~M.}\ \bibnamefont {Schmidt}},\
  }\href@noop {} {\bibfield  {journal} {\bibinfo  {journal} {Phys. Rev. B}\
  }\textbf {\bibinfo {volume} {93}},\ \bibinfo {pages} {116101(R)} (\bibinfo
  {year} {2016})}\BibitemShut {NoStop}%
\end{thebibliography}%

\end{document}